\documentclass[10pt,a4paper]{article}
\usepackage[utf8]{inputenc}
\usepackage{graphicx}
\usepackage[left=2cm,right=3cm,top=2cm,bottom=2cm]{geometry}
\usepackage{amsmath,amsfonts,amssymb}
\usepackage{pdfpages}

\usepackage{array,multirow,makecell}
\setcellgapes{1pt}
\makegapedcells
\newcolumntype{R}[1]{>{\raggedleft\arraybackslash }b{#1}}
\newcolumntype{L}[1]{>{\raggedright\arraybackslash }b{#1}}
\newcolumntype{C}[1]{>{\centering\arraybackslash }b{#1}}
\begin{document}
\title{\textbf{A wide diversity of 3D surfaces Generator using implicit functions}}
\maketitle
\begin{center}
\author{$Jelloul$  $EL MESBAHI_{1}$ \\
$ Ahmed ERRAMI_{2}$\\
$ Mohammed$  $ KHALDOUN_{2} $\\
$ Omar$ $BOUATTANE_{3}$\\
1 University Hassan II of Casablanca Faculty of Sciences, Morocco\\
2 Hassan II University of Casablanca , ENSEM,  Morocco\\
3 Hassan II University of Casablanca, ENSET, Morocco\\
  Email: j.elmesbahi@ensem.ac.ma} 
\end{center}
\section*{ABSTRACT}
In this paper we present an enhanced version of the paper [8]. It deals with a new family of implicit functions used to provide a wide diversity of 3D surfaces. This family involves some usual functions such as: the rectangular pulses, the saw-tooth pulses, the triangular pulses, the staircase function and the power function. By combining these usual functions, named constituent functions, in one implicit function,where some controlled parameters are included, we can easily insure a large  shape variability and deformation of the resulted 3D surfaces\\
 
\section{INTRODUCTION}
Today, there exists in the literature a multitude of works for synthesizing 3D surfaces. Most of theseworks involve approaches based either on procedural models or on the so-called surface models.We can also cite approaches based on chaos theory [1, 2].One technique is based on the principle described in [2],where , the representation of such a dynamic gives rise to mathematical objects called strange attractors.\\
The procedural models operate recursive procedures that allow a gradual growth of synthesized forms. The theory of fractals is probably the most exploited in the literature [3, 4]. As for surface models, they use more or less complex equations but not recursive. One example of such techniques using implicit functions defined by an equation:  F (x, y, z) = 0,   [5,6].\\
 As a surface model example, we can cite the well known such as the Klein bottle, Clebsh surface, Schwartz, surface and Chmutov surface. In this paper we will rely on this implicit function approch to generate a wide variety of 3D surfaces.
\section{CONSTITUENT  FUNCTIONS }
 In this paper we present an approach based on an implicit function. This latter is implemented using the following usual functions\\
\subsection{\textit{Curve representation by a piecewise model}}
The piecewise model of any  curve $ \mathcal{C} $  is generally viewed as a concatenation of a set of n curve pieces \\
$ Ci.\; \;     \mathcal{C} = CAT(Ci) (i=1 \;to\; n);  $
\\
Each piece Ci, delimited by its two extreme points  $Pi (x_{i},y_{i} )$   and   $P_{i+1} (x_{i+1},y_{i+1})$,  can be represented by any mathematical expression.\\
In this part we present a mathematical expression  of  the curve representation by a piecewise model of segments. Each piece Ci is represented by a segment Si  and the curve $\mathcal{C}$  is expressed by   concatenation equivalent function   $ f_{cc} (x) $    as: 
\begin{equation}
f_{{{\it cc}}} \left( x \right) =  \frac{1}{2}\,\sum _{i=1}^{n}   \frac{\left( 1-{\frac 
{ \left( x-x_{{i}} \right)  \left( x-x_{{i+1}} \right) }{ \left| x-x_{
{i}} \right|  \left| x-x_{{i+1}} \right| }} \right)  \left(  \left( y_
{{i+1}}-y_{{i}} \right) x+y_{{i}}x_{{i+1}}-y_{{i+1}}x_{{i}} \right) }{ \left( x_{{i+1}}-x_{{i}} \right) }
\end{equation}
The following expression  illustrates an example of the curve having eight  segments  using the  fcc model, figure 1.1.shows its resulted curve .
 
\begin{center}
 $ f_{{{\it cc}}} \left( x \right) = 0.5\,\left( f_{{{\it cc1}}} \left( x
 \right) + f_{{{\it cc2}}} \left( x \right) \right)  $ 
\end{center}
Where : \\
$ f_{{{\it cc1}}} \left( x \right) = \left( 1-{\frac { \left( x-1
 \right)  \left( x-2 \right) }{ \left| x-1 \right|  \left| x-2
 \right| }} \right)  \left( x-1 \right) + \left( 1-{\frac { \left( x-2
 \right)  \left( x-3 \right) }{ \left| x-2 \right|  \left| x-3
 \right| }} \right)  \left( 2\,x-3 \right) + \left( 1-{\frac { \left( 
x-3 \right)  \left( x-4 \right) }{ \left| x-3 \right|  \left| x-4
 \right| }} \right)  \left( 3\,x-6 \right) + \left( 1-{\frac { \left( 
x-4 \right)  \left( x-5 \right) }{ \left| x-4 \right|  \left| x-5
 \right| }} \right)  \left( -4\,x+22 \right) 
 $\\
 and\\
 $ f_{{{\it cc2}}} \left( x \right) = \left( 1-{\frac { \left( x-5
 \right)  \left( x-7 \right) }{ \left| x-5 \right|  \left| x-7
 \right| }} \right)  \left( \frac{1}{2}\,x+9/2 \right) +6-{\frac { \left( x-7
 \right)  \left( x-9 \right) }{ \left| x-7 \right|  \left| x-9
 \right| }}+ \left( 1-{\frac { \left( x-9 \right)  \left( x-10
 \right) }{ \left| x-9 \right|  \left| x-10 \right| }} \right) 
 \left( -x+10 \right) -5\,{\frac { \left( x-10 \right)  \left( x- 10.5
 \right) }{ \left| x-10 \right|  \left| x- 10.5 \right| }}
 $

\begin{center}
\includegraphics[scale=0.625]{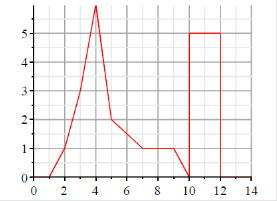} 
\end{center} 
\begin{center}
\textit{Figure 1.1. Example of a curve having eight piecewise segments }
\end{center}
Notice that we can easily transform each segment of fcc into a part of other curves. Figure 1.1. is made of eight segments. So, one can be replaced by any other expression.\\

\subsection{ EXAMPLES OF CONSTITUENT  FUNCTIONS }
\subsubsection{\textit{The saw- tooth pulse serie}}  It is  represented by the function :
\begin{equation}
f_{{st1}} \left( x \right) =    \frac{1}{2r}\,\sum _{i=-p}^{p} \left( 1-{\frac {
 \left( x-r \left( i-1 \right)  \right)  \left( x-ri \right) }{
 \left| x-r \left( i-1 \right)  \right|  \left| x-ri \right| }}
 \right)  \left( x-r \left( i-1 \right)  \right) 
\end{equation}

           Where    \qquad \qquad $  x\in[- \left( p+1 \right).r,p.r]  $ \\
r is the base of the right triangles  of saw-tooth  serie.
 \subsubsection{\textit{The triangular pulse serie:}}
 The serie of triangular pulse function is represented by :
 \begin{equation}
f_{{tr1}} \left( x \right) =\left| \sum _{i=-p}^{p}\frac{1}{r}\, \left( 1-{
\frac { \left( x-r \left( i-1 \right)  \right)  \left( x-ri \right) }{
 \left| x-r \left( i-1 \right)  \right|  \left| x-ri \right| }}
 \right)  \left( x-r \left( i-1 \right)  \right) - 1
 \right|
 \end{equation}
 Where :          $  x\in[- \left( p+1 \right).r,p.r]  $ 
  \quad and r is the base of the isoscele of each triangle pulse 
\subsubsection{\textit{Staircase serie:}}
 The staircase function is given by the  expression
 
\begin{equation}
f_{{sc1}} \left( x \right) = \frac{h}{r} \left( x-    \frac{1}{2}\,\sum _{i=-p}^{p} \left( 1-{
\frac { \left( x-r \left( i-1 \right)  \right)  \left( x-ri \right) }{
 \left| x-r \left( i-1 \right)  \right|  \left| x-ri \right| }}
 \right)  \left( x-r \left( i-1 \right)  \right)  \right) 
\end{equation}

 Where : \quad$  x \in [- \left( p+1 \right).r,p.r]  $  \qquad h = height of a stair level\qquad  and \qquad r = width of a bearing 
\subsubsection{\textit{Rectangular pulses function:}}
  The rectangular pulse series function can be represented by the following  different expressions :\\
 \begin{equation}
 f_{{rec1}} \left( x \right) =\frac{1}{2}-\frac{1}{2}\,{\frac {\sin \left( \pi \,x
 \right) +t}{ \left| \sin \left( \pi \,x \right) +t \right| }}\quad,\quad \left| t \right| <1
\end{equation}
\\
\begin{equation}
f_{{rec2}} \left( x \right) ={\frac {t}{r}}-\frac{1}{\pi}{\it arccot} \left( \cot
 \left( {\frac {\pi \, \left( x+t \right) }{r}} \right)  \right) +\frac{1}{\pi}{\it arccot} \left( \cot \left( {\frac {\pi \,x}{r}} \right)
 \right)  \quad,\quad  \frac{t}{r} \,\in \mathbb{R} \setminus \mathbb{Z}
\end{equation}

\begin{equation}
f_{{rec3}} \left( x \right) =   \frac{1}{2}\,\left( {\frac {\arcsin \left( \sin \left( \pi \,x \right)  \right) }{\arctan \left( \tan \left( \pi \,x \right)  \right) }}\,.\,{\frac {\arctan \left( \tan \left( \pi \, \left( x+t \right)  \right)  \right) }{\arcsin \left( \sin \left( \pi \, \left( x+t \right)  \right)  \right) }}\,-1\right)  \quad, \quad t \,\in \mathbb{R} \setminus \mathbb{Z}
\end{equation}
\subsubsection{\textit{The power function:}}
%  $f_{{p}} \left( x \right) =\sum _{i=0}^{N}a_{{i}}{x}^{b_{{i}}}{y}^{c_{{i}}}{z}^{d_{{i}}}$     \qquad  \qquad     
% where  \qquad $  a_{{i}},b_{{i}}, c_{{i}}$   and  $  d_{{i}}  \,\in \mathbb{R}     $

\begin{equation}
  f_{{p}} \left( x \right) =\sum _{i=0}^{N}a_{{i}}{x}^{b_{{i}}}{y}^{c_{{i}}}{z}^{d_{{i}}}     \quad,  \quad     
 where  \qquad   a_{{i}},b_{{i}},c_{{i}}\,   and\: d_{{i}}  \,\in \mathbb{R}     
\end{equation}

The constituent  functions can be represented by other usual expression as:
\subsubsection{\textit{The saw- tooth pulse function:}}
\begin{equation}
f_{{sr2}}\left( x\right) =\frac{1}{\pi}{\it arccot} \left( \cot \left( {\frac {\pi \,x}{r}}
 \right)  \right)
\end{equation}

 Where :   \qquad \qquad          r is the base of the right triangle 
 \subsubsection{\textit{The triangular pulse function:}}
 \begin{equation}
 f_{{tr2}}\left( x\right) =\frac{1}{\pi}\arccos \left( \cos \left( 2\,{\frac {\pi \,x}{r}} \right) 
 \right)
 \end{equation}

  \begin{equation}
 f_{{tr3}}\left( x\right)= \left| \frac{2}{\pi}\,{\it arccot} \left( \cot \left( {\frac {\pi \,x}{r
}} \right)  \right) -1 \right| 
 \end{equation}
 Where :         
  \quad   \quad r is the base of  isoscele  triangle 
\subsubsection{\textit{Staircase function:}}
 \begin{equation}
 f_{{st2}} \left( x \right) =h \left( {\frac {x}{r}}-  \frac{1}{\pi}{\it arccot}
 \left( \cot \left( {\frac {\pi \,x}{r}} \right)  \right) 
 \right)
 \end{equation}

 Where : \quad  \qquad h = height of a stair level\qquad  and \qquad r = width of a bearing 
%%%%%%%%%%%%%%%Exemple tableau%%%%%%%%%%%%%%%%%%%%%%%%%%%%%%%%%%%%%%%%%%%%%%%%%%%%%%
%\end{tabular}
%%%%%%%%%%%%%%%%%%%%%%%%%%%%%%%%%%%%%%%%%%%%%%%%%%%%%%%%%%%%%%%%%%%%%%%%%%%%%%%%%%%%%%%
\begin{center}
\begin{tabular}{|C{6cm}|C{6cm}|}
\hline \includegraphics[scale=0.625]{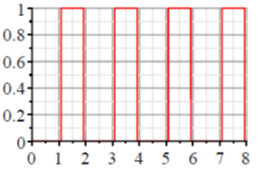} & \includegraphics[scale=0.625]{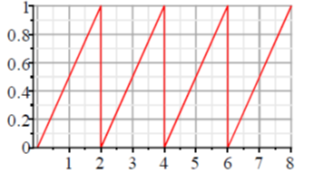} \\
\hline  \textit{Graphic  of equation(5) (6) and (7)}   & \textit{ Graphic  of equation (2) and (9)}  \\
\hline  \includegraphics[scale=0.625]{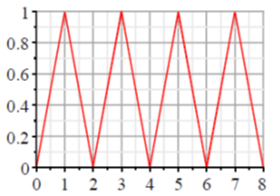}  & \includegraphics[scale=0.625]{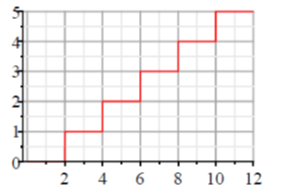} \\
\hline \textit{ Graphic  of equation(3)(10) and (11)}   & \textit{Graphic  of equation(4) and (12)}   \\
\hline 
\end{tabular}
\end{center}
\begin{center}
\textit{Figure 1.2. Graph of the above usual functions using Maple software.}
\end{center}
We emphasize in this paper that, only expressions containing trigonometric functions were used. By combining these constituent functions, we obtain an implicit equation that allows us to synthesize a wide variety of 3D surfaces. On the other hand, by making changes in  selecting constitutive functions at their parameters, we can further enrich the higher level of  the diversity of the  synthesized 3 D surfaces.
\section{FUNDAMENTAL   TERMS}
The implicit function built from a part of the constituent functions,  described in section 2, is as follows:

\begin{equation}
F \left( x,y,z \right) =\sum _{i=1}^{i} \left( \prod _{m=1}^{M} \left(  \left( {\frac {A
 \left( x,y,z \right) }{\mbox {D} \left( {x,y,z} \right) }}.{\frac {B
 \left( x,y,z \right) }{E \left( x,y,z \right) }} \right) .{\frac {C
 \left( x,y,z \right) }{F \left( x,y,z \right) }} \right)  \right) -d =0
\end{equation}
where \\
\begin{equation}
A \left( x,y,z \right) =\left( {{\it P1}_{{m,i}}}\right) ^{{\lambda1}_{{m,i}}} \hspace*{0.5cm}  and  \hspace*{0.5cm}  D \left( {x,y,z} \right) =  \left( {{\it P2}_{{m,i}}}\right) ^{{\lambda2}_{{m,i}}}
\end{equation}
\begin{equation}
B \left( x,y,z \right) = \left( H_{{t}} \left( {{\it P3}_{{m,i}}}^{{\lambda3}_{{m,i}}} \right)  \right) ^{{\mu1}_{m,i}}
and  \hspace*{0.5cm} 
 E \left( x,y,z \right) = \left( H_{{t}} \left( {{\it P4}_{{m,i}}}^{{\lambda4}_{{m,i}}} \right)  \right) ^{{\mu2}_{m,i}}
\end{equation}

\begin{equation}
 C \left( x,y,z \right) = \left( G_{{t}} \left( H_{{t}} \left( {\frac {
{{\it P5}_{{m,i}}}^{{\lambda5}_{{m,i}}}}{{{\it P6}_{{m,i}}}^{{\lambda6}_{{m,i}}}}} \right)  \right)  \right) ^{{\mu3}_{{m,i}}} 
 and  \hspace*{0.5cm} 
  F \left( x,y,z \right) = \left( G_{{t}} \left( H_{{t}} \left( {\frac {
{{\it P7}_{{m,i}}}^{{\lambda7}_{{m,i}}}}{{{\it P8}_{{m,i}}}^{{\lambda8}_{{m,i}}}}} \right)  \right)  \right) ^{{\mu4}_{{m,i}}} 
\end{equation}\\
and\\

P1, P2, P3, P4, P5, P6, P7, P8 are the  power functions,\\
d is a real constant, $\textbf{ H}_{{t}}  $  and $ \textbf{G}_{{t}} $  each represent either a direct trigonometric function and an inverse trigonometric function.\\
The variables and parameters used in the equations (14) to (16) are :  \\
  \begin{center}
 $   x \in [x_{{\min}},x_{{\max}}]$,\quad $ y \in [y_{{\min}},y_{{\max}}]$,\quad $ z \in [z_{{\min}},z_{{\max}}] $ \\
 $\lambda1,\,\lambda2,\,\lambda3,\,\lambda4,\,\lambda5,\,\lambda6,\,\lambda7,\,\lambda8,\,\mu1,\,\mu2,\,\mu3, \,\mu4 \,\in \mathbb{R } $    
 \end{center}
In the 3D  space, the  solutions of equation   \textbf{13} belong to a working space defined by the following:\\
Grid $ \left[  N_{x},\, N_{y},\,N_{z}\right] $ :\\
\hspace*{0.5cm} $ N_{x} $ is the number of points in the interval \quad  $ [x_{{\min}},x_{{\max}}]$ \\
\hspace*{0.5cm} $ N_{y} $ is the number of points in the interval \quad  $ [y_{{\min}},y_{{\max}}]$ \\
\hspace*{0.5cm} $ N_{z} $ is the number of points in the interval \quad  $ [z_{{\min}},z_{{\max}}]$ \\
\\
It is important to notify that each element \textbf{A}, \textbf{B}, \textbf{C}, \textbf{D}, \textbf{E} and \textbf{F} in expressions  \textbf{14} to \textbf{16} can be replaced by its absolute value.
\section{SENSITIVITY TO SETTING PARAMETERS CHANGES}
The most important feature of our approach is the level of sensitivity according to  the changes in parameters related to our proposed  implicit function.\\
Indeed, by varying some parameters, we produce  a very large variation in the obtained 3 D surface, while the variation of other parameters causes only a small change in these 3 D surface.
\subsection*{\textit{a - High sensitivity to parameters  variations:}}
Small changes in some parameters   produce a large variation in the resulting 3D shape.
In this case there exist two types of variations that will be expressed by giving two examples:\\\\
\textbf{1}-Taking the parameter \textbf{ m= 0.861} in the following equation A, we obtain the 3D shape $  \textbf{I}_{1}$     of  figure 2, if is \textbf{ m} is \textbf{0.851} value, we obtain another  3 D shape $ \textbf{I}_{2} $  , both shapes are significantly different, see Figure 2.
\\
\textbf{2}- in the equation B, using  the boundary values of x, y and z equal to 7 and a grid size of (34, 34, 34)  we  obtain the 3D shape $ \textbf{I}_{3} $ see Figure 3.
\\\\
In the  same equation B, using  the  boundary values of x,y and z equal to 7.05 and a same grid size of  (34, 34, 34) we  obtain the 3D shape $ \textbf{I}_{4} $  see Figure 3.
\\\\
In the  same equation B,   using  the boundary values of x, y and z equal to 7.05 and a grid size of (33, 33, 33)we  obtain  the 3D shape $ \textbf{I}_{5} $ of Figure 3.
\\\\
The shapes $ \textbf{I}_{3} $, $ \textbf{I}_{4} $ and $ \textbf{I}_{5} $, are significantly different. In some cases a small change in a parameter makes the shape disappear, this means that there is no solution  for implicite function.\\

%\subsubsection*{{4.a.1.example 1}}

\subsubsection*{4.a.1  Example 1:  Equation A }             
  $$ \left( \left(  \left| x \right|  \right) ^{ 3.3}+ \left(  \left| y \right| 
 \right) ^{ 3.3}+ \left(  \left| z \right|  \right) ^{ 3.3}-600 \right) +
 \left( A_{{x}}+A_{{y}}+A_{{z}} \right) ^{3}- 0.51\, \left( B_{{x}}+B_
{{y}}+B_{{z}} \right) ^{3}+100$$  
Where : 
$$ A_{{x}}= \left( {\it atan} \left( \tan \left( 1- \left(  \left| x
 \right|  \right) ^{ 0.31}+ 5.76\,{\frac { \left(  \left| x \right| 
 \right) ^{ 1.38}}{ \left|  \left(  \left| x \right|  \right) ^{ 1.3}+
 \left(  \left| y \right|  \right) ^{ 1.3}+ \left(  \left| z \right| 
 \right) ^{ 1.3} \right| }} \right)  \right)  \right) ^{3}$$
$$ A_{{y}}= \left( {\it atan} \left( \tan \left( 1- \left(  \left| y
 \right|  \right) ^{ 0.31}+ 5.76\,{\frac { \left(  \left| y \right| 
 \right) ^{ 1.38}}{ \left|  \left(  \left| x \right|  \right) ^{ 1.3}+
 \left(  \left| y \right|  \right) ^{ 1.3}+ \left(  \left| z \right| 
 \right) ^{ 1.3} \right| }} \right)  \right)  \right) ^{3} $$
 
$$ A_{{z}}= \left( {\it atan} \left( \tan \left( 1- \left(  \left| z
 \right|  \right) ^{ 0.31}+ 5.76\,{\frac { \left(  \left| z \right| 
 \right) ^{ 1.38}}{ \left|  \left(  \left| x \right|  \right) ^{ 1.3}+
 \left(  \left| y \right|  \right) ^{ 1.3}+ \left(  \left| z \right| 
 \right) ^{ 1.3} \right| }} \right)  \right)  \right) ^{3}$$
\\
$$ B_{{x}}= \left( {\frac { \left(  \left| x \right|  \right) ^{ 0.3}}{
 \left| \cot \left( mx \right)  \right| }} \right) ^{ 0.3} $$

$$ B_{{y}}= \left( {\frac { \left(  \left| y \right|  \right) ^{ 0.3}}{
 \left| \cot \left( my \right)  \right| }} \right) ^{ 0.3}$$
 
 $$ B_{{z}}= \left( {\frac { \left(  \left| z \right|  \right) ^{ 0.3}}{
 \left| \cot \left( mz \right)  \right| }} \right) ^{ 0.3}$$

\begin{center}
$   x\,\in\,[-155,155]$ , \quad $   y\,\in\,[-155,155]$ , \quad   $ z\,\in\,[-155,155]$ , \quad 
Grid $\left[82,82,82\right]$
\end{center}

\begin{center}
%%%%%%%%%%%%%%Exemple tableau%%%%%%%%%%%%%%%%%%%%%%%%%%%%%%%%%%%%%%%%%%%%%%%%%%%%%%
\begin{tabular}{C{5cm}C{5cm}}
 \includegraphics[scale=0.75]{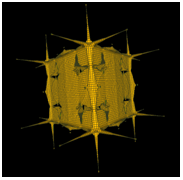} &\includegraphics[scale=0.72]{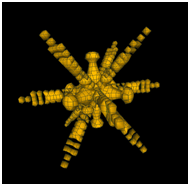}   \\
%\textbf{ I1: m=0.851 }& \textbf{I2 : m=0.861}  \\

\textbf{ $ \textbf{I}_{1} $: m=0.851 }& \textbf{$ \textbf{I}_{2} $ : m=0.861}  \\

\end{tabular}
%%%%%%%%%%%%%%%%%%%%%%%%%%%%%%%%%%%%%%%%%%%%%%%%%%%%%%%%%%%%%%%%%%%%%%%%%%%%%%%%%%%%%%
\end{center}

\begin{center}
\textit{Figure 2: Change in shapes  $ \textbf{I}_{1} $  to  $ \textbf{I}_{2} $  form  by changing parameters of the equation A}
\end{center}
\subsubsection*{4.a.2  Example 2:  Equation B  }
 
  $$ \alpha + \left(  \beta_{x} + \beta_{y}+ \beta_{z}\right) ^{3} -0.135 $$  
 Where :  
$$ \alpha=\frac{1}{24.4} \, \left( {x}^{2}+{y}^{2}+{z}^{2}-5 \right)  \left( 
 \left(  \left| x \right|  \right) ^{ 0.13} \left(  \left| y \right| 
 \right) ^{ 0.13} \left(  \left| z \right|  \right) ^{ 0.13}-5\right)  $$
$$\beta_{x}=\left| \cos \left( 2\,x \right)  \right|  \left( {\it atan} \left( 
\tan \left( 12\,{\frac {{x}^{2}+ \left(  \left| y \right|  \right) ^{2
}+{z}^{2}}{  {x}^{2}{y}^{2}+{z}^{2} }} \right)  \right) 
 \right) ^{2} \left( \cos \left(  0.1\,y \right)  \right) ^{-1}  $$
\\
$$\beta_{y}=\left| \cos \left( 2\,y \right)  \right|  \left( {\it atan} \left( 
\tan \left( 12\,{\frac {{y}^{2}+ \left(  \left| z \right|  \right) ^{2
}+{x}^{2}}{  {x}^{2}+{y}^{2}{z}^{2}  }} \right)  \right) 
 \right) ^{2} \left( \cos \left(  0.1\,y \right)  \right) ^{-1}  $$
\\
$$\beta_{z}=\left| \cos \left( 2\,z \right)  \right|  \left( {\it atan} \left( 
\tan \left( 12\,{\frac {{z}^{2}+ \left(  \left| x \right|  \right) ^{2
}+{y}^{2}}{  {y}^{2}+{x}^{2}{z}^{2} }} \right)  \right) 
 \right) ^{2} \left( \cos \left(  0.1\,y \right)  \right) ^{-1}$$ 
\begin{center}
$   x\,\in\,[-7,7]$ , \quad $   y\,\in\,[-7,7]$ , \quad   $ z\,\in\,[-7,7]$ , \quad 
Grid $\left[34,34,34\right]$ \quad for $ \left( \textbf{I}_{3}\right)  $
\end{center}
\begin{center}
$   x\,\in\,[-7.05,7.05]$ , \quad $   y\,\in\,[-7.05,7.05]$ , \quad   $ z\,\in\,[-7.05,7.05]$ , \quad 
Grid $\left[34,34,34\right]$ \quad for $ \left( \textbf{I}_{4}\right)  $
\end{center}
\begin{center}
$   x\,\in\,[-7.05,7.05]$ , \quad $   y\,\in\,[-7.05,7.05]$ , \quad   $ z\,\in\,[-7.05,7.05]$ , \quad 
Grid $\left[33,33,33\right]$ \quad for $ \left( \textbf{I}_{5}\right)  $
\end{center}

\begin{center}
\includegraphics[scale=0.70]{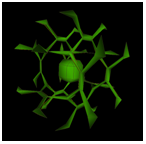}\qquad\qquad
\includegraphics[scale=0.72]{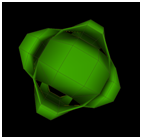} \qquad\qquad
\includegraphics[scale=0.72]{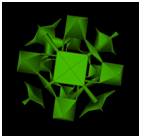}
\end{center}
 \hspace*{3cm}$ \textbf{I}_{3} $  \hspace*{4.5cm} $ \textbf{I}_{4} $  \hspace*{4.5cm} $ \textbf{I}_{5} $
\\
\begin{center}
\textit{Figure 3: The  shapes $ \textbf{I}_{3} $, $ \textbf{I}_{4} $,$ \textbf{I}_{5} $ are derived from the same  equation  B. In the case of $ \textbf{I}_{3} $  , the boundary of the variables x, y and z is 7  in a grid (34,34,34).  For  $ \textbf{I}_{4} $ , limits  of the variables x, y and z is 7.05 with a grid (34,34,34).For  $ \textbf{I}_{5} $  we have  the same limits as in the case of  but the grid is (33,33,33)}
\end{center}
\subsection*{\textit{b- Low sensitivity to parameter variations}}
In this case, a small change will produce a small change in the resulting 3D shape. The deformations may be obtained by modifying one or more parameters. As in the equation C, we propose the  parameter  changes of  m from  0.25  to 1 in steps of  0.25
\subsubsection*{ Example :  Equation C  }
 $$ F \left( x,y,z \right) =     \left(  \frac{\left(  \left| \cos \left(  0.7\,{m}{x}^{-1}
 \right)  \right|  \right)}{\left( \cos \left(  0.003\,{x}^{2}
 \right)  \right)}\right) ^{10} +{\frac { \left(  \left| \sin \left(  0.7\,my
 \right)  \right|  \right) ^{10} \left(  \left| \sin \left(  0.7\,mz
 \right)  \right|  \right) ^{10}}{ \left( \cos \left(  0.003\,{y}^{2}
 \right)  \right) ^{10} \left( \cos \left(  0.003\,{z}^{2} \right) 
 \right) ^{10}}}- 0.02\,\left( {x}^{3}+{y}^{2}+{z}^{2}\right) 
 $$  
 
\includegraphics[scale=0.71]{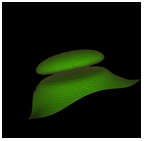}\quad
\includegraphics[scale=0.70]{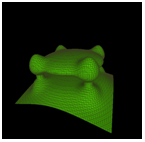} \quad
\includegraphics[scale=0.72]{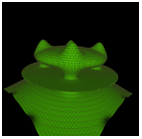}\quad
\includegraphics[scale=0.72]{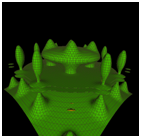}\\
 \hspace*{1.5cm}$ m=0.25 $  \hspace*{2.75cm}$ m=0.5 $  \hspace*{2.4cm} $ m=0.75 $ \hspace*{2.4cm} $ m=1$ 
\begin{center}
 \textit{ Figure 4.Modification obtained by varying the parameter m from   \textbf{0.25} to \textbf{1} with a step \textbf{0.25}}
\end{center}
\section{RESULTS}
To show the effectiveness of our model of implicit function, we propose another set of samples  equation examples to complete the first proposed set of examples in [8], using the visualization software \textbf{k3surf}, available at:\textbf{ http: // k3dsurf.sourceforge.net/index fr. Html.} 
Notice that, the use of another visualization software leads to slightly different shapes. 
\\
Notice that , the rectangular pulse  function  defined in subsection \textbf{ 2.2.4} can be integrated in a parametric function  as :
\begin{equation}
 X={\frac {f \left( x,y \right) }{ \left| f \left( x,y \right) 
\right| }}
\end{equation}
  
\begin{equation}
Y={\frac {g \left( x,y \right) }{ \left| g \left( x,y \right) 
\right| }}
\end{equation}
  
\begin{equation}
Z={\frac {h \left( x,y \right) }{ \left| h \left( x,y \right) 
\right| }}
\end{equation}
  
to provide the cubic  shapes (see the examples in the ancillary file “updatefile.pdf”,  generated by the MAPLE software). \\
\textbf{Notice that, this paper is associated to an ancillary file named "updatefile.pdf" which contains various images numbered from : 536 to 1109. these images represent the new results of this update version.}
 \section{CONCLUSION}
In this paper, we introduced a new family of implicit function ,  it is built from a set of common functions: rectangular pulses, triangular saw-tooth, the staircase function and power function. This new family of functions enabled us to generate a wide variety of 3D surfaces, and to set their deformations.\\
In the future work we will firstly, take a deep analysis of parameter changes effects on the deformation generated on the 3D shapes, and, we intend to implement new equations carried out by multiplying for example the exponents     $\lambda_{n}$ and $\mu_{n}$  in equations \textbf{14 }to \textbf{16} by the sine function, cosine, the rectangular series function, saw teeth and the staircase function\\
As a fundamental issue from this work, we developed an electronic device that was patented [9]. It automatically synthesizes more 3D surfaces than what we have presented in this paper.By this device the number of 3D surfaces produced in [7],[8] and by this version, remains smaller than the production capacity of this model.
 \section{ANNEXE}
\includepdf[pages = {1-130}]{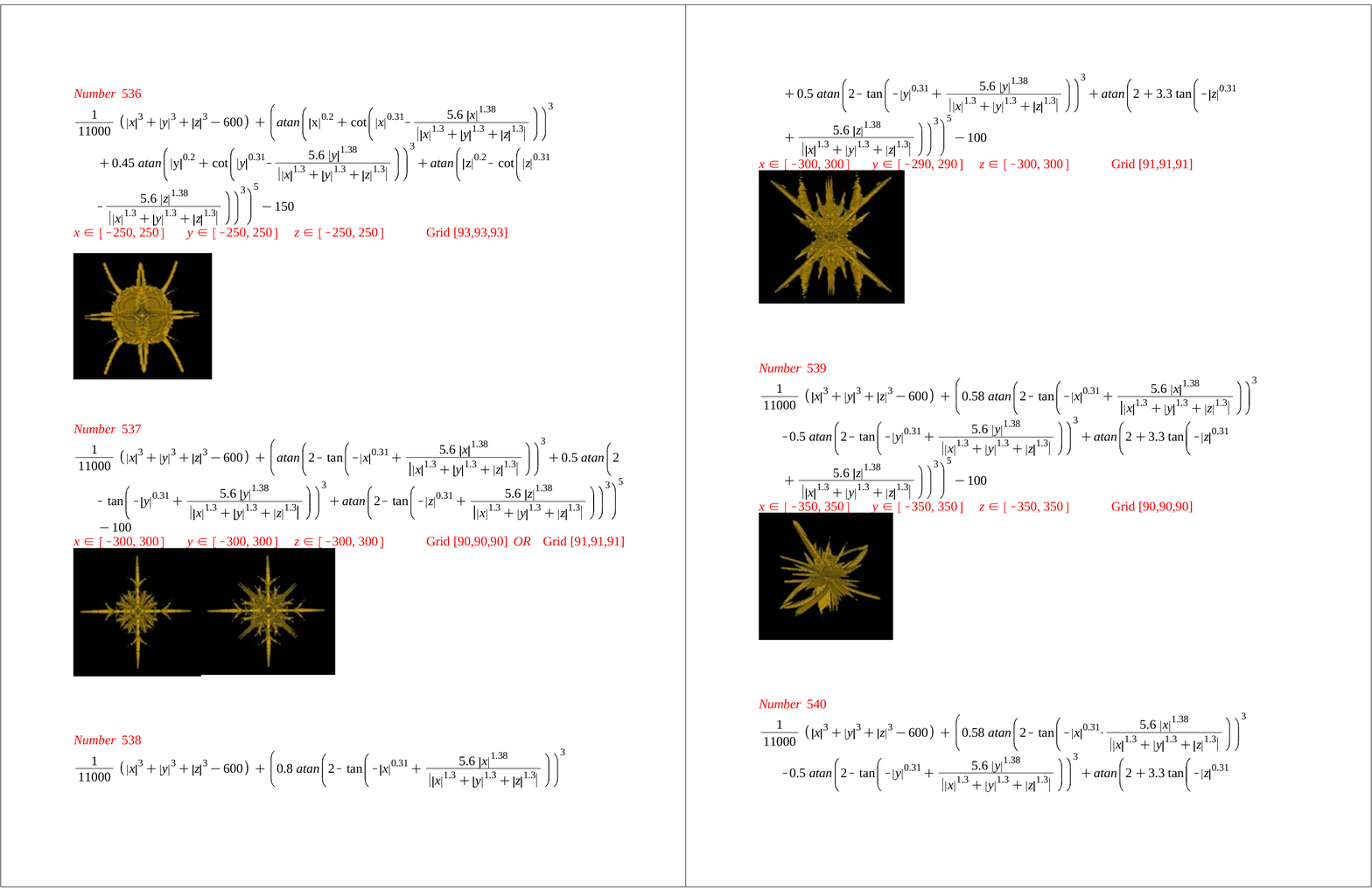}
\subsection*{\textit{REFERENCES}}

$[1] $ M. Field, M. Golubitsky , 'Symetry and chaos'. A search  for pattern in Mathematics, Art and Nature., second edition, SIAM. Philadelphia 2009.\\
\\
$[2]$   Safieddine Bouali et jos Leys, 'sculture du chaos', images des `mathématiques`, CNRS 2010.\\
\\
$[3] $ Mandelbrot, `Benoît` B. 'Fractal aspects of iteration if $ z \rightarrow Az(1-z)  $ for complex $ A $  and $ z $ . Annals of the New York Academy of Sciences. 1980, ss. 249-259.\\
\\
$[4]$  White, Daniel "The Unravelling of the Real 3D Mandelbulb "  On line 2009 on  \\
http://www.skytopia.com/project/fractal/mandelbulb.html.\\
\\
$[5]$    Blinn, J. F., "A Generalisation of Algebraic Surface Drawing", ACM Trans. Graphics, Vol. 1, No 3, July 1982, pp 135-256.\\
\\
$[6]$    J. Bloomenthal, C. Bajaj, J. Blinn, M.-P. Cani-Gascuel, A. Rockwood, B.Wyvill and G.Wyvill, 1997. Introduction to Implicit  Surfaces. Morgan Kaufmann.\\
\\
$[7]$  Jelloul Elmesbahi "Mathematical art gallery"  on line January 2013, on :  \\
https://plus.google.com/102117215461698921489/posts.\\
\\
$[8]$  Jelloul EL MESBAHI, Ahmed ERRAMI, Mohammed KHALDOUN and Omar BOUATTANE " A wide diversity of 3D surfaces Generator using a new implicit function".\\
 Arxiv1505.03977 $  on line on:  $ http://arxiv.org/find/cs/1/au:+Elmesbahi\_J/0/1/0/all/0/1\\
 \\
 $ [9 ] $ J. EL MESBAHI  , A. ERRAMI, M. KHALDOUN O. BOUATTANE " Electronic System for Synthesising 3D surfaces" .  WO2015084135 (A1). 
\end{document}